 \documentclass[12pt,preprint]{aastex}

\shorttitle{Direct $\ell$-Type Transitions in CRL\,618}
\shortauthors{Thorwirth et al.}

\begin{document}


\slugcomment{Version \today}
\title{Detection of HCN Direct $\ell$-Type Transitions Probing Hot 
Molecular Gas in the Proto-Planetary Nebula CRL\,618}


\author{S.~Thorwirth \altaffilmark{1},
F.~Wyrowski \altaffilmark{2,3},
P.~Schilke \altaffilmark{3}, K.~M.~Menten, \altaffilmark{3}
S.~Br\"unken \altaffilmark{1}, H.~S.~P.~M\"uller \altaffilmark{1}, and G.~Winnewisser \altaffilmark{1} }

\altaffiltext{1} {I. Physikalisches Institut, Universit\"at zu K\"oln,
Z\"ulpicher Str. 77, 50937 K\"oln, Germany\email{sven, bruenken, 
hspm, winnewisser@ph1.uni-koeln.de}}

\altaffiltext{2} {Department of Astronomy, University of Maryland, 
College Park, MD 20742-2421, USA}

\altaffiltext{3} {Max-Planck-Institut f\"ur Radioastronomie, Bonn, Germany
\email{wyrowski, schilke, kmenten@mpifr-bonn.mpg.de}}



\begin{abstract}

  We report the detection of direct $\ell$-type transitions towards the
  proto-planetary nebula CRL\ 618 during a study of vibrationally
  excited carbon chains.  The $J=8,10,11,12,13,14$ $\Delta J=0$ transitions
  of HCN in its first excited bending mode $v_2=1$ were detected in
  absorption against the continuum of the central  H\,{\sc ii} region making use of
  the Effelsberg 100\,m telescope and the Very Large Array.  Additionally, the $J=9$ direct
  $\ell$-type transition was detected in emission presumably indicating a
  weak maser. All lines are blueshifted with respect to the
  systemic velocity of CRL\,618 indicating that the lines originate from a
  hot, expanding circumstellar envelope. The HCN column density along the
  line of sight in front of the continuum is $2\times 10^{18}$\,cm$^{-2}$.
\end{abstract}


\keywords{line: identification --- radiation mechanisms: non-thermal - radio
  lines: stars --- stars: individual (CRL618) --- stars: AGB and post-AGB
  --- stars: mass loss and circumstellar matter}


\section{Introduction}

Molecules in vibrationally excited states are unique tools to study hot
molecular gas in interstellar and circumstellar environments.  The
vibrationally excited states can be pumped by IR radiation in the dusty
environment close to the exciting object (see e.g.\ \citealt{schilke92b,
schilke2000} or \citealt{wyrowski99} and references therein).

\begin{figure}
\plotone{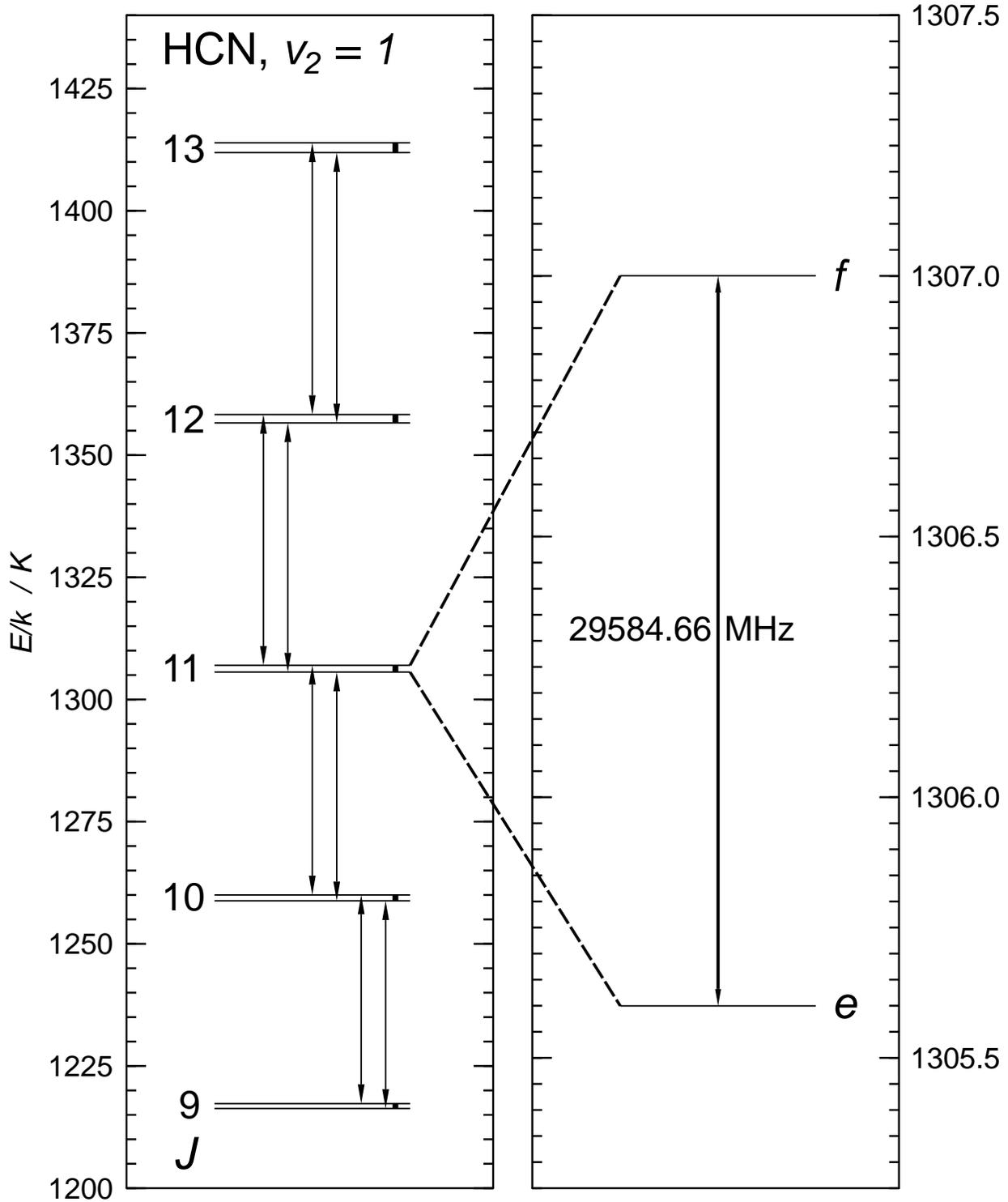}
\caption{Term value diagram of HCN in its $v_2=1$ vibrational state
  from $J=9$ to $J=13$.  On the left hand side both possible types of
  transitions with $\Delta J=\pm 1$ and the direct $\ell$-type transitions
  with $\Delta J=0$ are shown. The diagram on the right hand side
  shows the $J=11$ direct $\ell$-type transition at 29584.66\,MHz in detail.
  According to the convention of \citet{bro75} the vibrational substates
  have been labeled $e$ (lower) and $f$ (upper).\label{term-hcn}}
\end{figure}

A linear molecule with $N$ atoms has $3N-5$ vibrational modes. For HCN, this
yields four vibrational modes\footnote{For the sake of clarity it should be
  recalled that $\nu$ is used to denote a certain vibrational mode or the
  frequency whereas $v$ denotes the vibrational quantum number.}, two of
which are stretching modes (CN stretch $\nu_1$, CH stretch $\nu_3$), and the
$\mathrm{H-C-N}$ bending mode ($\nu_2$) which is doubly degenerate since the
molecule is free to bend in two orthogonal planes.  If the molecule is
bending and rotating simultaneously, the degeneracy is lifted giving rise to
a phenomenon denoted {\em $\ell$-type doubling}: The degenerate bending
state can be regarded as having components of an additional angular momentum
$p=\ell \hbar$ about the figure axis with $\ell=v,v-2,v-4,...,-v$.
For a first excited bending mode
$\ell=+1,-1$ causes the splitting of every rotational level into two
sublevels. Now two different types of transitions with either $\Delta J=\pm
1$ or $\Delta J=0$ may occur, the latter being denoted {\em direct
$\ell$-type transitions} (cf. Figure \ref{term-hcn}).
Since the splitting
of the sublevels of a given $J$ is rather small, the corresponding
transitions occur at low frequencies. For a molecule in a first excited
bending mode those are to first order given by $\nu=qJ(J+1)$, where $q$ is
the $\ell$-type doubling constant ($q_{\mathrm{(HCN},v_2=1)}\approx
224$\,MHz).

In the laboratory, direct $\ell$-type transitions were first measured by
\citet{shu50} for the molecules HCN and OCS. In space, rotational
transitions of vibrationally excited HCN ($\Delta J=\pm 1$, $v_2=1,2$;
``vibrational satellites'') were first detected by \citet{ziu86} toward
Orion-KL and IRC+10216. Meanwhile, HCN was also detected in much higher
vibrationally excited states (e.g. $v_2=4$,
\citealt{schilke2000,schilke2002}). A detection of the $J=19$ direct
$\ell$-type transition of HCN ($v_2=1$) in IRC+10216 has previously been
reported by \citet{cer96}, although no details are given.

We started a detailed study of vibrationally excited carbon chain molecules
toward the protoplanetary nebula CRL\,618 at radio and millimeter
wavelengths to examine their radial distribution by measuring their
vibrational temperatures. In this \textit{letter}, we report the first
detection of seven consecutive direct $\ell$-type transitions of HCN ($v_2=1$) at cm
wavelengths, commencing with $J=8$ through 14. The results on the higher
cyanopolyynes HC$_3$N and HC$_5$N
are reported separately \citep{wyr2002,thorwirth-phd}.

\section{Observations}

\subsection{Effelsberg 100\,m observations}

The observations were performed during 8 observing sessions from 1999 March
to 2002 January employing the 100\,m radiotelescope of the
Max-Planck-Institut f\"ur Radioastronomie at Effelsberg, Germany.  As
frontends the 0.65\,cm, 1\,cm, 1.3\,cm, and 1.9\,cm HEMT receivers with
typical receiver temperatures from 30 to 70\,K, respectively, were used.
The FWHM beam widths and the frequencies of the HCN lines are given in Table
\ref{analysis}. Pointing was checked every one to two hours on CRL\,618
itself or other appropriate radio sources resulting in an average pointing
accuracy of 4\arcsec .  The intensity scale was established using continuum
drift scans on
W3(OH), NGC\,7027 and 3C147 and comparing with the flux densities given by 
\cite{ott94} . Pointing scans on CRL\,618 were used to determine
its flux density relative to these calibration sources (Table \ref{analysis}).

\subsection{CSO observations}

Spectra of the $J=4-3$ line of HCN in its $\nu_2$ bending mode were obtained
in 1999 December with the 10.4 m telescope of the Caltech Submillimeter 
Observatory (CSO)\footnote{The CSO is operated by the California Institute
  of Technology under funding from the National Science Foundation, Grant
  No. AST-9980846.} with a
system temperature of 600~K. The line was observed in the upper sideband and
several different observing frequencies were used to avoid blending from
lines in the lower sideband. The spectrometer and observing procedure
are described by \citet{menten95}. The CSO beamsize at 356 GHz is 26\arcsec\ and
we assumed a beam efficiency of 78\% (taken from the CSO webpage).

\subsection{VLA observations}

CRL 618 was observed in the $J=13$ direct $\ell$-type transition (cf. Table
\ref{analysis}) with the Very Large Array (VLA\footnote{The VLA is operated
  by the National Astronomy Observatory, a facility of the National Science
  Foundation operated under cooperative agreement by Associated
  Universities, Inc.}) in its B configuration, leading to an angular
resolution of 0.12\,$^{\prime\prime}$. At the time of the observations, 15
antennas were equipped with 0.7\,cm receivers. A total bandwidth of 6.25~MHz
was observed with 128 channels and the spectral resolution was 48.8 kHz. The
total time on source was 3.5\,h. Regular observations of 0555+398, 3C48 and
3C84 were used for amplitude, flux and bandpass calibration, respectively.
The phase was selfcalibrated on the strong continuum emission from CRL\,618.

The remaining non Q-band antennas were used to observe the $J=4$ direct
$\ell$-type transition at 4488.48\,MHz. No line was detected at an RMS noise
level of 4 mJy in 0.8 km/s wide channels.  The total continuum flux at this
frequency is $26\pm 3$~mJy.

\section{Results and Discussion}

\begin{figure}
\epsscale{0.55}
\plotone{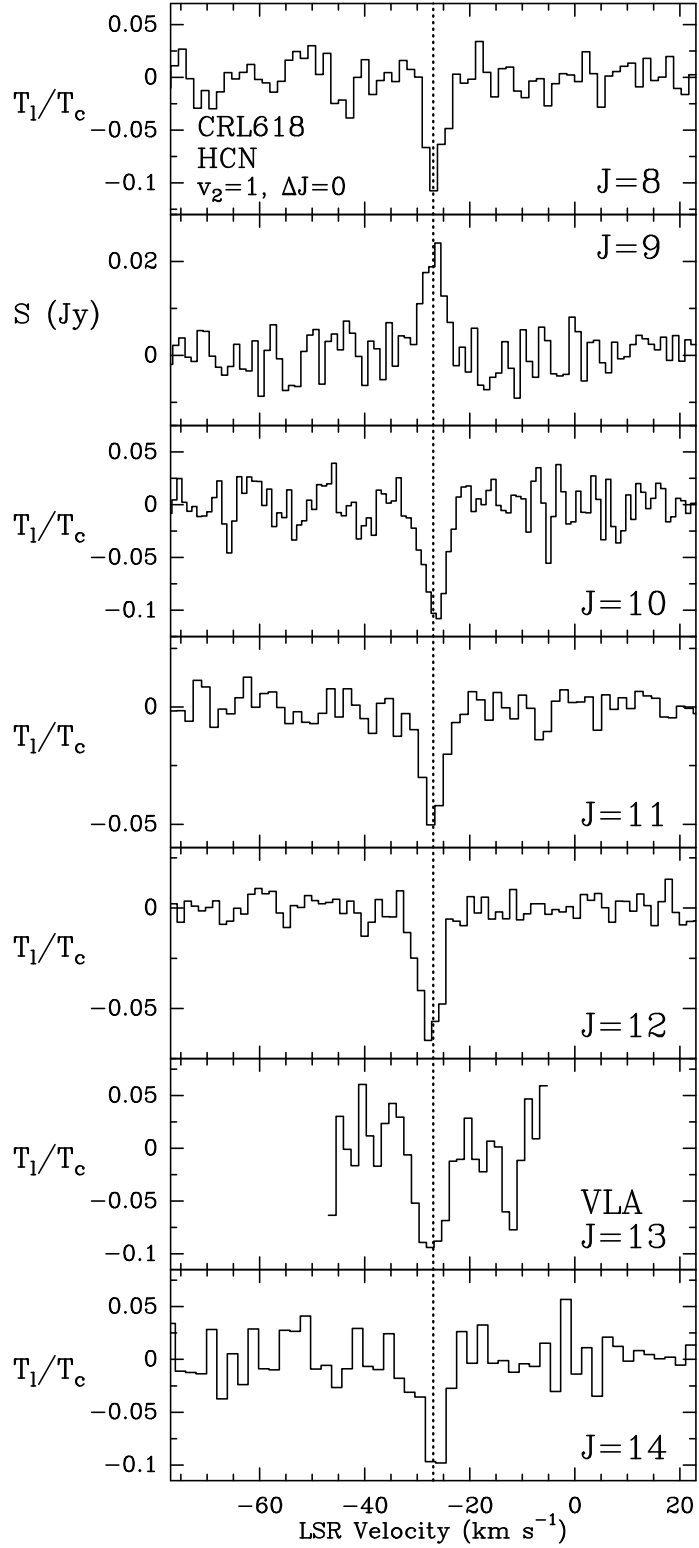}
\caption{The $J=8,9,10,11,12,13,14$ $\Delta J=0$ (direct $\ell$-type) transitions
  detected toward CRL\,618. The $J=8,9,10,11,12,14$ transitions were detected
  using the 100m telescope and the $J=13$ transitions using the VLA,
  respectively. The dashed line marks the line velocity
  $v_{L}= -27$\,km\,s$^{-1}$ indicating a significant blueshift from
  $v_{sys}=-24.2$\,km\,s$^{-1}$ of CRL\,618.
\label{lines_final}}
\end{figure}

\begin{figure}
\epsscale{0.95}
\plotone{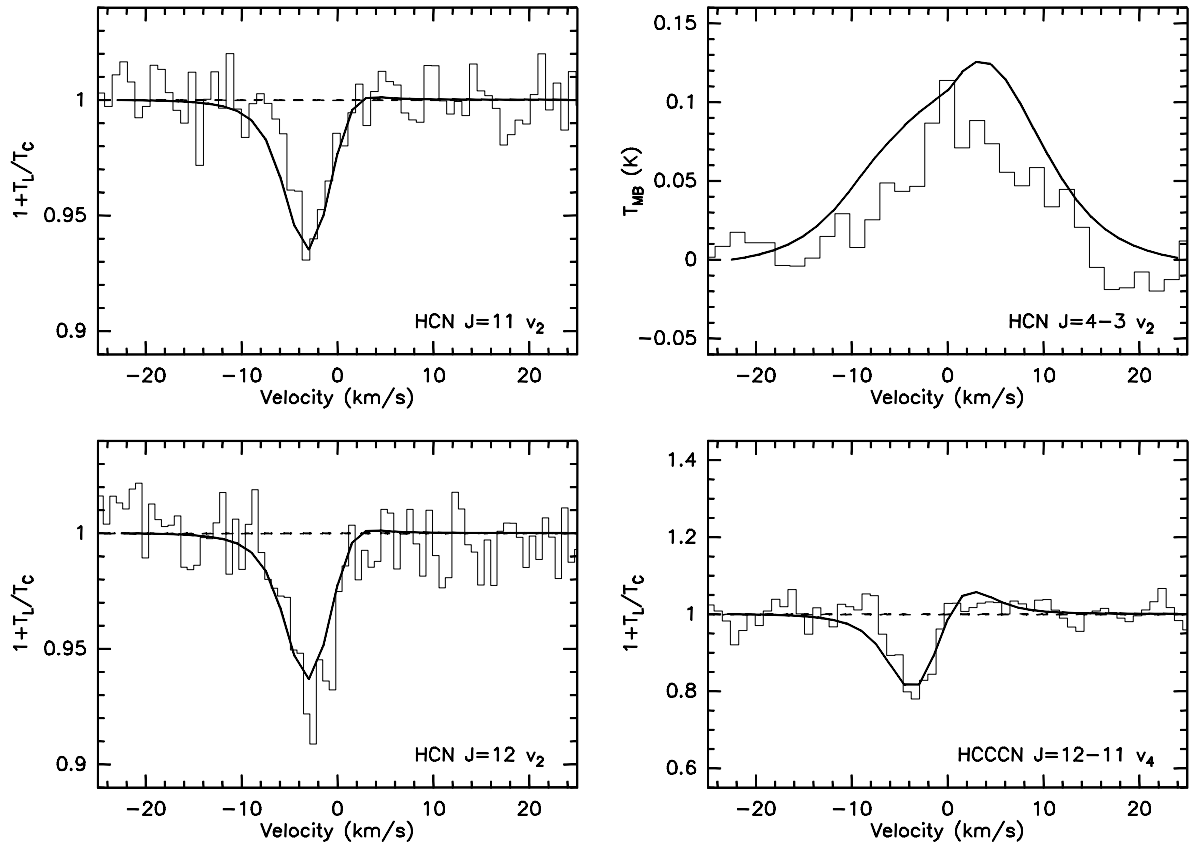}
\caption{Comparison of observed spectra of vibrationally excited
  HCN/HC$_3$N with results of an expanding envelope model.}
  \label{3d_model}
\end{figure}

Figure \ref{lines_final} shows the direct $\ell$-type lines observed toward
CRL\,618 and the line parameters from Gaussian fits to the spectra are given
in Table \ref{analysis}. The $J=8,10,11,12,13,14$ direct $\ell$-type
transitions appear as absorption lines towards the continuum of the H\,{\sc
  ii} region with line velocities of approximately $v_L=-27$\,km\,s$^{-1}$.
The systemic velocity $v_{sys}$ of CRL\,618 is $-$24.2\,km\,s$^{-1}$
\citep{wyr2002}, hence all of the observed lines are blueshifed relative to
$v_{sys}$ indicating that the lines originate from a hot, expanding
circumstellar envelope.
To determine the physical conditions of the absorbing gas, we used a
spherical LTE model of an expanding envelope, developed to interpret our
observations of vibrationally excited HC$_3$N derived by \citep{wyr2002}.
To fit the
HCN lines we use the temperature, density and velocity structure of the
expanding envelope, which fits the HC$_3$N lines, and only vary the HCN
abundance. Using a temperature of 560~K (see discussion below)
a HCN column density of $2\times 10^{18}$~cm$^{-2}$ is
needed to cause the observed absorption.
The resulting fits to the spectra are shown 
in Figure \ref{3d_model} together with a spectrum of the HC$_3$N $v_4=1$
$J=12-11$ transition which has a similar upper energy ($\sim 1300$\,K)  
Since no continuum flux measurements were
performed at 350~GHz, the HCN $J=4-3$ spectra are shown in brightness
temperature units, whereas for the other spectra the ratio of line to
continuum temperature, which reduces calibration uncertainties, is shown.
The deviation of the HCN $J=4-3$ model from the observed spectrum could be
due to pointing and/or focus errors: a pointing error of 10\arcsec\ alone
would explain the difference between observation and model and cannot be
excluded. The model consists of power laws for temperature, density and
velocity starting at an inner radius of 0.11\arcsec\, and an H\,{\sc ii}
region within that radius, which fits the continuum measurements of CRL~618.
The temperature at the inner radius is 560~K.  In the model, the HCN $J=4-3$
line is highly optically thick and mostly sensitive to the model temperature
and the emitting size. The HCN direct $\ell$-type lines, on the other hand,
are optically thin and probe a combination of temperature and column density
of the model. The best fit model has a HCN/HC$_3$N abundance ratio of 3 to 6
dependent on the assumed population of the vibrational levels of HC$_3$N
which is consistent with the result obtained by mid-infrared absorption
measurements by \citet{cer2001}.

\begin{figure}
\epsscale{0.95}
\plotone{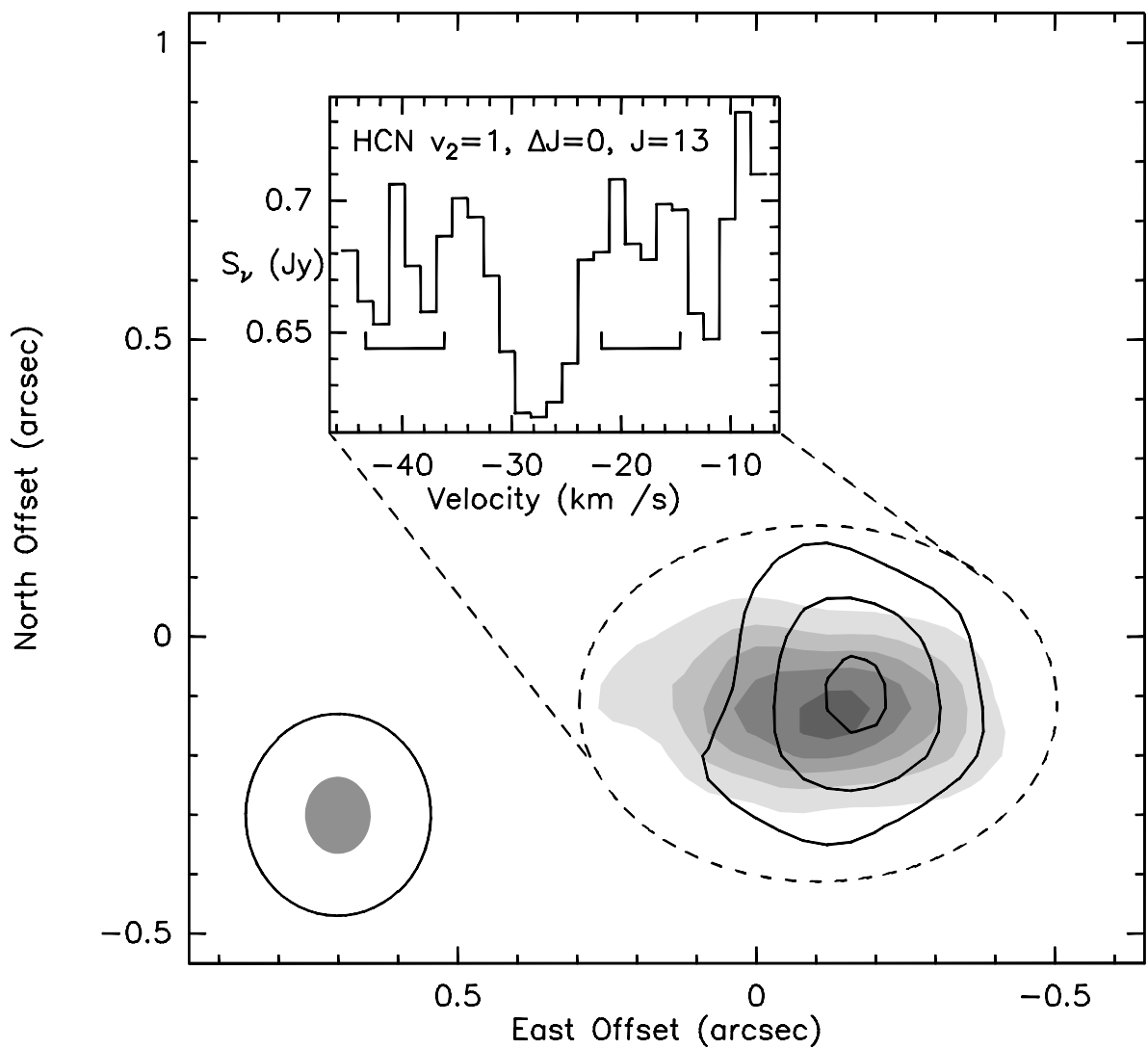}  
\caption{Results of the Q-band VLA observations. The continuum emission
  is shown in greyscale with steps of 10, 30, 50, 70, and 90\% of the peak
  continuum flux (140~mJy/beam). The HCN absorption is shown in contours of
  $-$16.5, $-$27.5, $-$38.5 mJy/beam, averaged from V$_{lsr}$ $-$24.7 to
  $-$30.3 km/s. The beamsizes are indicated in the lower left and are 0.13
  by 0.11\arcsec and 0.34 by 0.31\arcsec, respectively. The insert in the
  upper left shows a spectrum of the emission integrated over the region
  within the dashed ellipse.
  \label{vla-hcn}}
\end{figure}

Figure \ref{vla-hcn} shows the results of the VLA observations. The total
continuum flux density at 40 GHz, estimated from the flux on the shortest
baselines, is 0.75 Jy with an uncertainty of 10\%. The size of the continuum
emission is $0.34\times 0.16$\arcsec, estimated from a Gaussian fit to the
UV data.  To increase the spectral sensitivity, every four channels were
averaged together and a taper in the UV plane was applied, reducing the
angular resolution to $0.34\times 0.31$\arcsec . The insert in Figure
\ref{vla-hcn} shows the spectrum integrated over the indicated area. Line
parameters of a Gaussian fit to the spectrum are given in Table
\ref{analysis}.
To image the HCN absorption the continuum was subtracted from the data using
the channel ranges marked in the insert of Figure \ref{vla-hcn}. The
contours in Figure \ref{vla-hcn} show the HCN absorption averaged over the
line.


The observed 40~GHz continuum emission compares well with the results of
\citet{martin-pintado93,martin-pintado95} at 23~GHz. The HCN absorption falls
into the same velocity range as the hot core (HC) component seen in ammonia
by \citet{mar92}. The HCN absorption is slightly
shifted to the west from the center of the continuum, which was also
observed for the HC ammonia component and the hot dense disk observed
by \citet{martin-pintado95}. However, no absorption is observed at
the ammonia broad absorption velocity of $-$50\,km\,$s^{-1}$ which
\citet{martin-pintado93,martin-pintado95} interpreted as occurring from
post-shocked clumps. Accordingly, the HCN absorption most likely originates
from the same volume of gas as the ammonia hot core component and the dense
disk.

\subsubsection*{A weak $J=9$ maser?}

In contrast to the $J=8,10,11,12,13,14$ transitions appearing as absorption
lines toward the continuum, the $J=9$ transition is found in emission.
Blending with an unknown line cannot be ruled out entirely, but to the best
of our knowledge there is no known transition of a different molecule, and
no trace of a $J=9$ absorption (which would modify the emission-profile of
the blending line) is seen.  Moreover, the (blueshifted) velocity and the
line width correspond well to the velocities and linewidths of the
$J=8,10,11,12,13,14$ transitions. In particular the linewidth argument is
compelling, since other emission lines are much broader (e.g.\ the linewidth
of the HC$_7$N $J=21-20$ transition is 27~km~s$^{-1}$, \citealp{mar92}).
These facts suggest that the $J=9$ transition is a weak maser amplifying the
continuum.  Its optical depth has a different sign, but is similar in value
to the absorption lines.

How can this maser be understood? A quantitative analysis would require
quite detailed modeling of the pumping mechanism, which is not feasible
because many of the transition rates (in particular collision rates between
vibrational states and within the vibrationally excited state) are not or
only poorly known.  However, one can argue as follows that this transition
is easily perturbed or inverted: The splitting due to $\ell$-type doubling
is much smaller than the rotational splitting.  Considering the rotational
level system $J=8$, $J=9$ and assuming that the excitation temperature of
the $J=9e-8e$ is $T_{\rm ex}^{9e,8e}$ (and equal to $T_{\rm ex}^{8f,8e}$),
while the excitation temperature for the $f$ (upper) states is taken to be
$T_{\rm ex}^{9f,8f} = T_{\rm ex}^{9e,8e} + \Delta T$ (with $\Delta T \ll
T_{\rm ex}^{9e,8e}$), one can show that inversion occurs ($T_{\rm
  ex}^{9f,9e} < 0$) if the following condition is met:
\begin{equation}
  \label{eq:tex}
  \Delta T > \frac{\nu_{9f,9e}}{\nu_{8f,8e}}  T_{\rm ex}^{9e,8e} =
  0.025\, T_{\rm ex}^{9e,8e} 
\end{equation}
In terms of occupation numbers, a difference $\Delta T$ = 2.5\% means that
for an excitation temperature of 560~K, the occupation of the $J=9f$ level
needs to be elevated in population by only 0.17\% with respect to a
completely thermalized distribution to invert the direct $\ell$-type
transition.  If one considers the populations relative to the vibrational
ground state, similar arguments can be made.  A similar system is the
ammonia molecule, where the inversion splitting is much smaller than the
rotational splitting, and indeed, some masers are found for ammonia as well
(see \citealt{swm} or \citealt{madden}). It has to be emphasized that the
actual pumping mechanism is probably quite different for the ammonia masers,
the similarity is that it is not difficult to produce masers in the
inversion lines.

Having argued that a small perturbation is sufficient, we still have to
identify a possible cause for perturbations in our $\ell$-type system.  One
good candidate is a line overlap.  Such a mechanism has been invoked
successfully for pumping of OH masers (e.g.\ \citealt{cm}), although the
details of the pumping mechanism are different.  One possible candidate
responsible for a perturbation in the $J=8,9$ $v_2=1$ levels is an overlap
of rotational lines of the vibrationally excited molecule: For the $J=8-7$
and $J=9-8$ transitions of the $v_2=1$ and $v_2=2$ states the $\ell$
components $1f$ and $2e$ are separated by only 11 and
58\,MHz\footnote{$J=8-7$, $v_2=2$, $\ell=2e$ at 712361\,MHz; $J=8-7$,
  $v_2=1$, $\ell=1f$ at 712372\,MHz; $J=9-8$, $v_2=2$, $\ell=2e$ at
  801421\,MHz; $J=9-8$, $v_2=1$, $\ell=1f$ at 801363\,MHz}, respectively,
corresponding to $\Delta v=5$\,km\,s$^{-1}$ and $\Delta v=22$\,km\,s$^{-1}$.
Accordingly, the first overlap could occur locally whereas the second one
could connect different parts of the envelope. A corresponding maser effect
has been observed for SiO masers by \citet{cernicharo1991,cernicharo1993}.
%
%


\subsubsection*{HCN Direct $\ell$-type transitions in High-Mass Star-Forming
Regions}

Initiated by the results presented here, additional searches for direct
$\ell$-type transitions of HCN have been performed towards high-mass star
forming regions. So far, we were able to
detect the $J=9$ transition toward Orion-KL and Sgr\,B2(N) and the $J=9,10$
transitions toward G10.47+0.03 using the Effelsberg 100\,m
telescope (Thorwirth et al., in preparation).
Direct $\ell$-type transitions are optically thin and hence
can be used to derive reliable column density estimates of the hot gas
component. Moreover, the VLA observations of CRL\,618 presented here
demonstrate that this special kind of transitions can be used to observe hot
gas at high angular resolution and low frequencies. 

\section{Summary}
Using the Effelsberg 100m telescope and the VLA, we have detected seven
direct $\ell$-type transitions of HCN in its $v_2=1$ state.
All lines appear in absorption against the embedded H\,{\sc II}
region, except for the $J=9$ transition, which shows weak maser action. The
observed line velocities agree well with those observed in vibrationally
excited HC$_3$N by Wyrowski et al. (2002), who used their extensive data
to model the physical parameters of the hot, dense emission region. VLA
observations of the $J=13$ line reveal the emission region is compact and
covers only the western part of the embedded H\,{\sc ii} region, similarly
to ammonia.  Since in addition to CRL 618 we detected $\ell$-type HCN lines
toward hot molecular cores in regions of high-mass star formation, they also
represent an interesting new tool to study the immediate vicinity of young
(proto-)stellar objects.

Triggered by the radioastronomical observations presented here, a laboratory
investigation of direct $\ell$-type transitions of HCN ($v_2=1$) has been
carried out in the Cologne laboratory covering rotational quantum numbers up
to $J=35$ at 278.7\,GHz (Thorwirth et al., in preparation) The complete
analysis will be presented in a following paper.

\acknowledgments

The present study was supported by the Deutsche Forschungsgemeinschaft (DFG)
via Grant SFB\,494 and by special funding from the Ministry of Science of
the Land Nordrhein-Westfalen.  FW is supported by the National Science
Foundation under Grant NSF AST-9981289. We would also like to thank an
anonymous referee for valuable suggestions on the maser interpretation.


\clearpage

\begin{deluxetable}{cccccccr@{}lr@{}l}
  \footnotesize \tablecaption{Direct $\ell$-type transitions of HCN (MHz) in
    its $v_2=1$ vibrational state observed in the present study, as well
    as corresponding upper energies (K), beam sizes $\theta_{mb}$
    ($^{\prime\prime}$), line to continuum ratios, line intensities $I_L$ (mJy), line
    velocities $v_L$ ($\mathrm{km\,s^{-1}}$), line widths $\Delta v$ 
    ($\mathrm{km\,s^{-1}}$) and continuum flux densities $S_{\nu}^{cont}$ (mJy).
  \label{analysis}}
\tablewidth{0pt}
\tablehead{
\colhead{$J$} & \colhead{Frequency\,\tablenotemark{a}} & \colhead{$E_u$}  &
$\theta_{mb}$ & $-T_L/T_C$&$I_L$ & $v_L$ & \multicolumn{2}{c}{$\Delta v$} & \multicolumn{2}{c}{$S_{\nu}^{cont}$} } 
\startdata
4  &  4488.48 & 1067 &      & \tablenotemark{b}  &      &            &    &        & 26\,&(3)\\
8  & 16148.55 & 1178 & 52   & \tablenotemark{c}  &      & $-$26.6\,(3) & 3.&8\,(5)   & \\
9  & 20181.40 & 1217 & 42   &             & 22\,(4) & $-$27.0\,(3)        & 4.&7\,(7)   &425\,&(60)\\
10 & 24660.31 & 1259 & 34   & 0.11\,(3)   &      & $-$26.6\,(3)        & 4.&4\,(6)   &490\,&(90)\\
11 & 29584.66 & 1306 & 28   & 0.05\,(1)   &      & $-$27.0\,(2)        & 4.&8\,(6)   &\multicolumn{2}{c}{\tablenotemark{d}}\\
12 & 34953.76 & 1358 & 24   & 0.07\,(2)   &      & $-$27.4\,(2)        & 5.&1\,(4)   &680\,&(140)\\
13 & 40766.90 & 1413 & 0.33 & 0.11\,(4) &      & $-$27.3\,(7)        & 5.&4\,(12)  &750\,&(75)\\
14 & 47023.20 & 1473 & 20   & 0.11\,(4)   &      & $-$26.7\,(4)        & 4.&9\,(11)  &840\,&(85)\\
\enddata
\tablenotetext{a}{Laboratory frequencies taken from \citet{maki74}.}
\tablenotetext{b}{No line detected.}
\tablenotetext{c}{Line detected in absorption without valid calibration.}
\tablenotetext{d}{No absolute flux calibrator available.}
\end{deluxetable}

\clearpage

\clearpage

\clearpage

\end{document}